\documentclass[aps,prb,reprint,showpacs,amsmath,superscriptaddress]{revtex4-1}
\usepackage{amssymb}
\usepackage{mathrsfs}
\usepackage{graphicx}
\usepackage{float}
\usepackage{txfonts}
\usepackage[normalem]{ulem}
\usepackage{color}
\usepackage{bm}

\begin{document}
\title{Engineering topological phases with a three-dimensional nodal-loop semimetal}
\author{Linhu Li}
\affiliation{Department of Physics, National University of Singapore, 117542, Singapore}

\author{Han Hoe Yap}
\affiliation{NUS Graduate School for Integrative Sciences and Engineering, Singapore 117456, Republic of Singapore}
\affiliation{Department of Physics, National University of Singapore, 117542, Singapore}

\author{Miguel A. N. Ara\'ujo}
\affiliation{Beijing Computational Science Research Center, Beijing 100089, China}
\affiliation{CeFEMA, Instituto Superior
T\'ecnico, Universidade de Lisboa, Av. Rovisco Pais, 1049-001 Lisboa,  Portugal}
\affiliation{Departamento de F\'{\i}sica,  Universidade de \'Evora, P-7000-671,
\'Evora, Portugal}

\author{Jiangbin Gong}
\email{phygj@nus.edu.sg}
\affiliation{Department of Physics, National University of Singapore, 117542, Singapore}
\affiliation{NUS Graduate School for Integrative Sciences and Engineering, Singapore 117456, Republic of Singapore}

\begin{abstract}
A three-dimensional (3D) nodal-loop semimetal phase is exploited to engineer a number of intriguing phases featuring different peculiar topological surface states. In particular, by introducing various two-dimensional gap terms to a 3D tight-binding model of a nodal-loop semimetal, we obtain a rich variety of topological phases of great interest to ongoing theoretical and experimental studies, including chiral insulator, degenerate-surface-loop insulator, second-order topological insulator, as well as Weyl semimetal with tunable Fermi arc profiles.  The unique concept underlying our approach is to engineer topological surface states that inherit their dispersion relations from a gap term.    The results provide one rather unified principle for the creation of novel topological phases and can guide the search for new topological materials. Two-terminal transport studies are also carried out to distinguish the engineered topological phases.
\end{abstract}

\maketitle
{\it Introduction}. The generation and realization of novel topological phases of matter continue to be one main research frontier
in condensed-matter physics.
Topologically nontrivial matter is often featured and manifested by their fascinating symmetry-protected surface states.  A topological insulator in $d$-dimension supports metallic states in {its} $(d-1)$-dimensional surface{s} \cite{Hasan_2010,Qi_2011,Bernevig_book,Shen_book}. In a topological semimetal, the valence and conduction bands touch in a region of co-dimension $p>1$. For example, in nodal-point semimetals {(NPSM)}, Fermi surfaces are zero-dimensional (0D) points connected by Fermi arcs in the surface Brilloun zone (BZ) \cite{Wan_2011,Young_2012,Morimoto_2014,Yang_2014}, whereas in the
three-dimensional (3D) nodal-line (or loop) semimetals (NLSM) \cite{Burkov_2011,Weng_2015,Yu_2015,Kim_2015,Zhang_2016,Yan_2016,Li2017},  the touching region of bulk bands is one or several one-dimensional (1D) loops, with degenerate surface states within the projection of the loop in the 2-dimensional (2D) surface BZ, termed as the ``drumhead'' surface states.
In addition, NLSM with a myriad of loops with different topological features has been proposed \cite{Fang_2015,Lim_2017,Zhong_2017,Ezawa_Hopf,Yan_Hopf,Ezawa_crossing,Yan_crossing,Li2017_2}. Experimental studies of NLSM have also been carried out with various methods recently \cite{Experiment1,Experiment2,Experiment3,Experiment4}.


 Perspectives linking different topological phases can be especially constructive and insightful. In particular, NLSM subject to a restricted class of gap terms opening up the bulk bands can lead to familiar topological insulator phases \cite{Li2016,Li2017}, with their topological feature characterized by a winding number defined for a gap term along the nodal loop of the original NLSM.  The same winding number can be connected with the number of Dirac cones in the surface BZ.
 We are thus motivated in this work to obtain a rich variety of topological phases under a unified scheme, namely, the introduction of an additional 2D gap term with a spin degree of freedom to 3D NLSM.  By simply tailoring the gap term whose dispersion relation can be carried over to the topological surface states,  a number of intriguing topological phases can be engineered,  including the previously discussed chiral insulator and those just constructed most recently by other means. For example, we obtain an insulating phase with degenerate surface loop (DSL) in certain parameter regimes, and a higher-order topological insulator with 1D surface states on the hinges of two surfaces \cite{Benalcazar2017,schindler2017,Langbehn2017,Imhof_2017}.
    Given that distinctively different topological surface states can be engineered by a tailored gap term applied to a common NLSM, studies of topological transitions between them will be fruitful.  To distinguish these phases, we computationally study their two-terminal transport properties and discuss their different behaviors in transport measurements.







{\it General approach.}  Consider a 3D tight-binding Hamiltonian
\begin{eqnarray}
H_{\rm NL}=\left(t\sum_{i=x,y,z}\cos{k_i}-m\right)\tau_1+\sin{k_z}\tau_3,
\end{eqnarray}
with $\tau$ the Pauli matrices acting on a pseudospin-1/2 space, e.g. a sublattice space. $t$ is a hopping parameter, which is scaled to unity and serves as the energy scale hereafter.
The dispersion relation of $H_{\rm NL}$ is given by $E_{\rm NL}=\sqrt{\left(\sum_{i=x,y,z}\cos{k_i}-m\right)^2+\sin^2{k_z}}$,
which yields one or two nodal loops.  Throughout, we focus on the case of $1<m<3$, with a single nodal loop $\mathbf{L}_0$ defined by $\cos k_x+\cos k_y=m-1$  in the $k_x$--$k_y$ plane with $k_z=0$.  This nodal loop is protected by ${\cal P}{\cal T}$ symmetry,
where ${\cal P}$ is the conventional parity operator and ${\cal T}$ represents the time reversal operator, with ${\cal P}{\cal T} H_{\rm NL}(\mathbf{k}){\cal T}^{-1}P^{-1}=H_{\rm NL}(\mathbf{k})$.  A $\mathbf{k}$-dependent gap term $M(\mathbf{k})\tau_2$, if turned on,  will break the ${\cal P}{\cal T}$ symmetry, yielding a Weyl semimetal or a trivial band insulator (BI).

To generate different topological phases rather systematically, we introduce the following gap term on top of $H_{\rm NL}$,
\begin{eqnarray}
H_{\rm gap}=\left[M_0(\mathbf{k})s_0+\sum_{i=1,2,3} M_i(\mathbf{k})s_i\right]\tau_2,\label{general_gap}
\end{eqnarray}
with $s_0$ the $2\times2$ identity matrix, $s_{i}$ the Pauli matrices acting on another (pseudo) spin-1/2 degree of freedom. The existence of surface states of $H=H_{\rm NL}+H_{\rm gap}$ (say along $z$), as well as their energy dispersion, is related to $H$ projected on a plane in the 3D BZ perpendicular to $k_z$ \cite{Mong_2011}.  Below we take advantage of this explicit
bulk-edge relation to facilitate the construction of surface states localized along the $z$ direction. To that end we consider a gap term which is independent from $k_z$,
so that $H_{\rm gap}$ is only a function of $\mathbf{k}_{\parallel}=(k_x,k_y)$. Physically, $H_{\rm gap}=H_{\rm gap}(\mathbf{k}_{\parallel})$ may describe a 2D spin-orbit coupling. The spectrum of $H_{\rm gap}(\mathbf{k}_{\parallel})$ itself is given by $\alpha M_{\rm gap,\pm}(\mathbf{k}_{\parallel})$, where $\alpha=\pm1$ is the eigenvalue of $\tau_2$, and
\begin{equation}
M_{\rm gap, \pm}(\mathbf{k}_{\parallel})= M_0(\mathbf{k_{\parallel}})\pm\sqrt{ \sum_iM_i^2(\mathbf{k_{\parallel}}}). \label{Meq} \end{equation}
The dispersion relation of the total Hamiltonian $H=H_{\rm NL}+H_{\rm gap}$ can be found accordingly, which is given by
\begin{eqnarray}
E={\pm}\sqrt{E_{\rm NL}^2+M_{\rm gap, \pm}^2(\mathbf{k}_{\parallel})}.\label{E_all}
\end{eqnarray}



The surface states of $H=H_{\rm NL}+H_{\rm gap}$ under OBC along the $z$ direction can now be analyzed from the effective Hamiltonian $
H_{\rm eff, \pm}=H_{\rm NL}+ M_{\rm gap, \pm}(\mathbf{k}_{\parallel})\tau_2$ as a $2\times 2$ Dirac Hamiltonian winding with $k_z$ \cite{Burkov_2011}.  Because $M_{\rm gap, \pm}(\mathbf{k}_{\parallel})$ is independent of $k_z$, this gap term does not affect the winding behavior of $H_{\rm eff, \pm}$, and hence will not play any role in determining the existence of 2D surface states. Indeed,
applying the method in Ref.~{\cite{Mong_2011,sup}} 
, we arrive at the following condition for surface states to exist, namely, 
\begin{eqnarray}
|\cos{k_x}+\cos{k_y}-m|<1. \label{condition_1}
\end{eqnarray}
Thus, surface states of $H$ can only appear in the region bounded by the nodal loop $\mathbf{L}_0$ associated with the original $H_{\rm NL}$.  Remarkably,  under this condition,  the dispersion relation of the surface states is found to be precisely $\pm M_{\rm gap, \pm}(\mathbf{k}_{\parallel})$ 
{\cite{sup}}.  That is, provided that surface states do exist, the 3D Hamiltonian $H$ directly inherits the dispersion relations of its surface states from $\pm M_{\rm gap, \pm}(\mathbf{k}_{\parallel})$, the spectrum of the gap term $H_{\rm gap}$ itself.  It is this finding that allows us to engineer below a variety of topological phases by tailoring their surface state dispersions at our will. From  Eq.~(\ref{E_all}), it is clear that when the zero-energy surface states overlap with $\mathbf{L}_0$, the 3D system becomes gapless.

{\it Chiral insulators.}  As a simple case, we discuss how a chiral topological insulator \cite{Hosur2010} can be created from our approach.  We set $M_0$ and one of the three $M_i$'s (say $M_3$) to be both zero \cite{Li2017}. $H=H_{\rm NL}+H_{\rm gap}$  then possesses a new ${\cal P}{\cal T}$ symmetry with ${\cal P}s_2  H^{*}(\mathbf{k}) s_2 {\cal P}^{-1}=H(\mathbf{k})$,
 as well as a chiral symmetry with $\mathcal{S} H(\mathbf{k}) \mathcal{S}^{-1}=-H(\mathbf{k})$, where $\mathcal{S}=s_3\tau_2$.
In this case, the two effective Hamiltonians $H_{\rm eff, \pm}=H_{\rm NL}+ M_{\rm gap, \pm}(\mathbf{k}_{\parallel})\tau_2$ yield identical spectrum, implying that $H$ is two-fold degenerate.
The surface states, if they exist, should be also two-fold degenerate. Consider a specific example with
\begin{eqnarray}
H_{\rm gap}=\left[(\sin{k_x}) s_2+(\sin{k_y}+\mu) s_1\right]\tau_2, \label{ex2}
\end{eqnarray}
where $M_0=M_3=0$. The spectrum of $H_{\rm gap}$ itself represents a 2D nodal-point semimetal.  When $\mu=0$, a {four-fold degenerate 2D Dirac point} at $(k_x,k_y)=(0,0)$ lies at the center of $\mathbf{L}_0$ in the 2D surface BZ, as shown in Fig.\ref{fig1}(a). The full system $H=H_{\rm NL}+H_{\rm gap}$ inherits this dispersion relation via its surface states.
As $\mu$ is tuned, the Dirac point moves around in the surface BZ.  When the Dirac point falls right on $\mathbf{L}_0$, the bulk gap closes and the system undergoes a topological phase transition.  Consistent with this, as the Dirac point leaves $\mathbf{L}_0$ [Fig.\ref{fig1}(b)], the system becomes a trivial insulator. {Due to the chiral symmetry $\mathcal{S}$, the system belongs to either AIII or DIII class, depending on whether time-reversal symmetry is respected.  In both cases, the topological properties are characterized by an invariant in $\mathbb{Z}$, e.g.  a winding number assigned to the nodal loop $\mathbf{L}_0$,
\begin{eqnarray}
\nu=\oint_{\mathbf{L}_0}\frac{M_1(\mathbf{k}_{\parallel}) dM_2(\mathbf{k}_{\parallel})-M_2(\mathbf{k}_{\parallel}) dM_1(\mathbf{k}_{\parallel})}{M_1^2(\mathbf{k}_{\parallel})+M_2^2(\mathbf{k}_{\parallel})}.\label{winding_n}
\end{eqnarray}
This winding number is equivalent to the Chern number defined on a 2D plane containing $\mathbf{L}_0$ \cite{Li2016}.

\begin{figure}
\includegraphics[width=0.8\linewidth]{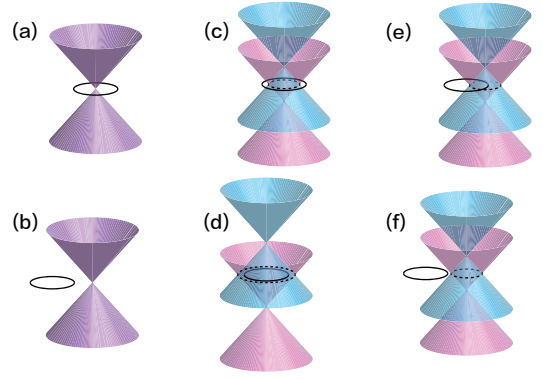}
\caption{Sketches of surface Dirac cones inherited by $H=H_{\rm NL}+H_{\rm gap}$ from $H_{\rm gap}$ and the nodal loop $\mathbf{L}_0$ (black solid lines) of $H_{\rm NL}$. Dashed lines represent the DSL obtained from the zero solution in Eq.~(\ref{Meq}). A DSL exists only if it lies within the region enclosed by $\mathbf{L}_0$. The Dirac cones in (a) and (b) are two-fold degenerate, whereas this degeneracy is lifted in (c)-(f) by a nonzero $m_0$ \cite{note}.}\label{fig1}
\end{figure}

{\it Degenerate surface loop insulator and Weyl semimetal.}
Next we introduce a {$\mathbf{k}$-independent $M_0(\mathbf{k})=m_0$}, which breaks both the ${\cal P}{\cal T}$ symmetry and the chiral symmetry mentioned above.
 A simple gap term of this type with $M_3=0$ is given by
\begin{eqnarray}
H_{\rm gap}=\left[m_0+(\sin{k_x}) s_2+(\sin{k_y}+\mu) s_1\right]\tau_2. \label{Hgap_SNL}
\end{eqnarray}
Without loss of generality, we choose $m_0\geqslant 0$ hereafter.
The two-fold degenerate spectrum of the chiral insulator is lifted, so is the four-fold degenerate surface Dirac point. Indeed, focusing on the spectrum of $H_{\rm gap}$, the mere effect of a nonzero but constant $M_0$ is to shift its Dirac cones by $\pm m_0$ in energy, which results in a gapless loop at zero energy as shown by the dashed lines in Fig. \ref{fig1}(c)-(f). According to our general insights above, this feature is now carried over to the surface states of $H$ and so we have
degenerate surface zero modes along a 1D loop.
Hence, by tuning the magnitude of $m_0$, one can engineer the {size} of the degenerate surface loop (DSL). We stress that the DSL is robust unless it falls outside of $\mathbf{L}_0$ in the surface BZ [Fig.\ref{fig1}(c) and (d)].  We also note in passing that DSL here is protected by the symmetry
\begin{equation}
\mathcal{C}H^*(\mathbf{k})\mathcal{C}^{-1}=-H(\mathbf{k}), \label{eq:DSLsymmetry}
\end{equation}
with $\mathcal{C}=s_1 \tau_2$. Such a symmetry is a combination of parity and particle-hole symmetries \cite{Schnyder_2008},} which is ensured by $M_3=0$, namely, the absence of $s_3 \tau_2$ in Eq.~(\ref{Hgap_SNL}).

Further, a nonzero $\mu$ in $H_{\rm gap}$ depicted in Eq.~(\ref{Hgap_SNL}) shifts the DSL center, yielding more topological phases [Fig.\ref{fig1}(e) and (f)] as we tune $\mu$.  Indeed, as the DSL center moves and touches $\mathbf{L}_0$, the spectrum of $H=H_{\rm NL}+H_{\rm gap}$ becomes gapless and a topological phase transition occurs.  
Specifically, one finds the following gap closing conditions,
\begin{eqnarray}
&&k_z=0,~~\cos{k_x}+\cos{k_y}-(m-1)=0,\nonumber\\
&&\sin^2{k_x}+(\sin{k_y}+\mu)^2= m_0^2.\label{phases}
\end{eqnarray}
 For $m=2$, $H$ describes an insulator with DSL under the condition $\mu+m_0<1$, and a trivial band insulator otherwise.

\begin{figure}
\includegraphics[width=0.7\linewidth]{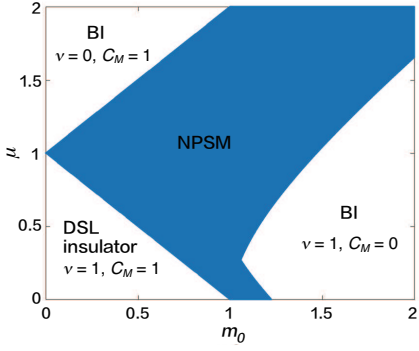}
\caption{A topological phase diagram with $m=2$, with the phase boundary determined from Eq.~(\ref{phases}). Different insulating phases are characterized by two topological invariants $\nu$ and $C_M$ defined in the main text, which are ill-defined in the semimetallic NPSM phase.}\label{fig2}
\end{figure}

Because the winding number $\nu$ defined in Eq. (\ref{winding_n}) is unrelated to $M_0(\mathbf{k}_{\parallel})=m_0$, and yet changing $m_0$ can close the bulk gap, we see that the topological invariant $\nu$ does not suffice to characterize different types of insulator phases with DSL.
To find a second topological invariant, we return to the effective Hamiltonians $
H_{\rm eff, \pm}=H_{\rm NL}+ M_{\rm gap, \pm}(\mathbf{k}_{\parallel})\tau_2$ defined earlier. The two effective Hamiltonians
$H_{\rm eff, \pm}$ depict a nodal-loop semimetal $H_{\rm NL}$ subject to a gap term $M_{\rm gap, +}(\mathbf{k}_{\parallel})\tau_2$ or
$M_{\rm gap, -}(\mathbf{k}_{\parallel})\tau_2$.  In either case, the surface loop due to $M_{\rm gap, +}(\mathbf{k}_{\parallel})\tau_2$ or
 $M_{\rm gap, -}(\mathbf{k}_{\parallel})\tau_2$ alone is not protected by a bulk gap because it can shrink to a point and disappear altogether without closing the bulk gap of $H_{\rm eff, \pm}$. However, our DSL phase must be described by the two effective Hamiltonians at the same time.
Indeed, what is protected by a nonzero bulk gap is the intersection of the two pairs of surface Dirac cones, given by $\alpha M_{\mathrm{gap},\pm}$ with $\alpha=\pm1$ respectively.
This motivates us to seek a topological invariant based on both surface Dirac cones.  Specifically,
consider the Berry curvature along the loop $\mathbf{L}_0$,
\begin{eqnarray}
\mathbf{V}_{\pm}^{\mathbf{L}_0}=\nabla_{\mathbf{k}}\times\mathbf{A}^{\mathbf{L}_0}_{\pm}{=} {\frac{\mathrm{sgn}(M^{\mathbf{L}_0}_{\mathrm{gap},\pm})}{4\pi|M^{\mathbf{L}_0}_{\mathrm{gap},\pm}|^2} }(\sin{k_y},-\sin{k_x},0),
\end{eqnarray}
with $\mathbf{A}_{\pm}^{\mathbf{L}_0}$ being the Berry connection on the lower band of Hamiltonians $H_{\mathrm{eff},\pm}$, evaluated along $\mathbf{L}_0$.
So long as the bulk is gapped, $M_{\mathrm{gap}, +}(\mathbf{k}_{\parallel})$ and $M_{\mathrm{gap}, -}(\mathbf{k}_{\parallel})$ must be nonzero along $\mathbf{L}_0$, and hence cannot change their respective signs. Thus, either the Berry curvatures
$\mathbf{V}_{+}^{\mathbf{L}_0}$ and $\mathbf{V}_{-}^{\mathbf{L}_0}$ always point in the same direction along the loop $\mathbf{L}_0$, or they always point in the opposite direction along the loop $\mathbf{L}_0$.
Such a binary feature can be captured by the integer
\begin{eqnarray}
C_{M}=[\mathrm{sgn}(M^{\mathbf{L}_0}_{\mathrm{gap},+})-\mathrm{sgn}(M^{\mathbf{L}_0}_{\mathrm{gap},-})]/2,
\end{eqnarray}
where $C_M=0$ for the first and $C_M=1$ for the latter case.
Consider now $m_0=0$ (which yields $C_M=1$). For each fixed $k_x,k_y$ on the nodal loop $\mathbf{L}_0$, $M^{\mathbf{L}_0}_{\mathrm{gap},-}$ increases with $m_0$, and changes sign after the DSL intersects $\mathbf{L}_0$ (and then the DSL vanishes as $m_0$ is further increased). This topological phase transition is captured by a jump in $C_M$ from $C_M=1$ to $C_M=0$.
On the other hand,  changing $\mu$ can move the DSL center out of the loop $\mathbf{L}_0$, thus turning the DSL insulator into a trivial BI without affecting $C_{M}$.  Thus both $\nu$ and $C_M$ are needed to characterize the topological properties of gapless surface states. A full phase diagram based on this insight as well as the conditions listed in Eq.~(\ref{phases}) is presented in Fig.~\ref{fig2}. In the (blue) shaded region of Fig.~\ref{fig2}, the DSL and $\mathbf{L}_0$ touch at discrete points, yielding a NPSM (Weyl) semimetal. That there is no fine-tuning needed for the NPSM phase can also be understood by inspecting the ways for $\mathbf{L}_0$ and the DSL to touch, which span a continuous parameter regime. The Fermi arcs connecting different Weyl points are given by the portion of the DSL that stays inside $\mathbf{L}_0$, as illustrated in Fig. \ref{fig1}(e).  Since the calculations of both $\nu$ and $C_M$ require a nonzero gap along $\mathbf{L}_0$, they are not defined in the NPSM phase.

\begin{figure}
\includegraphics[width=1.0\linewidth]{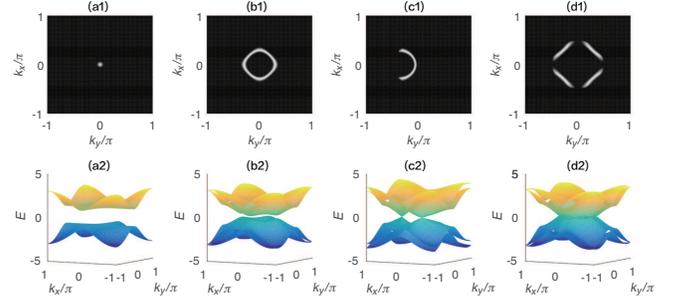}
\caption{The surface zero modes with $m=2$, and their corresponding bulk spectra with $k_z=0$ in (a2)-(d2). The zero modes are indicated by the bright {markers} in (a1)-(d1). The parameters are (a) $m_0=0$, $\mu=0$; (b) $m_0=0.8$, $\mu=0$; (b) $m_0=0.8$, $\mu=1$; (d) $m_0=1.1$, $\mu=0$.}\label{fig3}
\end{figure}

In Fig.~\ref{fig3} we present the zero-energy modes with open boundary condition (OBC) along the $z$ direction, and their corresponding bulk spectra. The zero-energy surface states form a single point, a closed ring, or one or several arcs in the surface BZ [Fig.~\ref{fig3}(a1)-(d1)], corresponding to a 3D chiral insulator, a DSL insulator, or a nodal point semimetal respectively. In Fig.~\ref{fig3}(d) there are four Fermi arcs connecting eight nodal points. This is because for $\mu=0$, although both the DSL and $\mathbf{L}_0$ are centered at $k_x=k_y=0$, neither of them is a perfect circle, and they cannot fully overlap. Instead, the DSL and  the loop $\mathbf{L}_0$ cross each other at multiple points in the momentum space.


{\it Second-order insulators.}  As the final topological phase created by our general approach,
  we consider a nonzero $M_3(\mathbf{k}_{\parallel})$.  In this case, $H_{\rm gap}$ itself in general depicts
  a 2D insulating system that can be topologically nontrivial.  The presence of the $s_3 \tau_2$ term in $H=H_{\rm NL}+ H_{\rm gap}$
  breaks the symmetry $\mathcal{C}$ previously discussed, and as a result the 2D surface states are fully gapped.
  As an example, we consider
\begin{eqnarray}
H_{\rm gap}=\left[\sin{k_x} s_2+\sin{k_y} s_1+ (\mu'-\cos{k_x}-\cos{k_y})s_3\right]\tau_2.\label{gap_HO4}\nonumber\\
\end{eqnarray}
$H_{\rm gap}$ above is similar to a 2D topological insulator model \cite{QWZ_2006}, with two topologically nontrivial phases for $-2<\mu'<0$ and $0<\mu'<2$, with a 2D Chern number $C=\pm1$.  For $|\mu'|>2$,  $H_{\rm gap}$ itself describes a 2D trivial BI.
Below we only consider positive values of $\mu'$.

\begin{figure}
\includegraphics[width=0.8\linewidth]{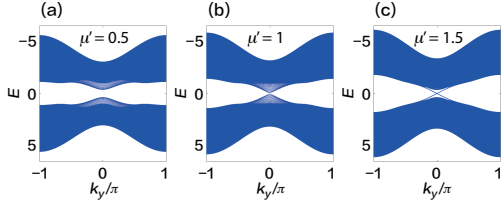}
\caption{Spectrum versus $k_y$ with open boundary condition in $x$ and $z$ directions, for the gap terms as in Eq. (\ref{gap_HO4}). The parameters are $m=2$, and $\mu'=0.5,1,1.5$ for the three panels, respectively.}\label{fig4}
\end{figure}

The surface states of $H$ under OBC along $z$, which inherit their dispersion relations from that of $H_{\rm gap}$,
 should be fully gapped in the 2D $k_x$--$k_y$ surface BZ. Because $H_{\rm gap}$ itself can be topologically nontrivial, it is anticipated that upon opening another direction, gapless edges can emerge.  This is confirmed in Fig.~\ref{fig4}(c), where we show the spectrum versus $k_y$, with OBCs in both $x$ and $z$. We observe gapless states localized in the 1D edges of a 3D sample along the $y$ direction for $1<\mu'<2$, thus realizing a second-order topological insulator \cite{Benalcazar2017,schindler2017,Langbehn2017,Imhof_2017}, a subject of great interest recently. Such intriguing ``edge states of surface states" are obtained here using exactly the same philosophy as how we engineer other interesting topological phases discussed above.


We now explain the criterion ($1<\mu'<2$) for the generation of a second-order topological insulator. Following \cite{Li2016}, we consider the gap term as a 2D nodal loop semimetal, $(\mu'-\cos{k_x}-\cos{k_y})s_3$, gapped by $\sin{k_x} s_2+\sin{k_y} s_1$. The topological property of $H_{\mathrm{gap}}$ can then be characterized by a chirality assigned to the 2D loop $\bm{l}_0:\cos{k_x}+\cos{k_y}=\mu'$. Hence, for this nontrivial topology to manifest as surface states of the 3D system $H$, the parent 3D loop $\mathbf{L}_0$ must be able to capture the chirality of $\bm{l}_0$, i.e. $\mathbf{L}_0$ must enclose $\bm{l}_0$, whence the condition $\mu'>1$. To understand the upper bound, it suffices to notice that, for $\mu'>0$, the gap term $H_{\mathrm{gap}}$ admits non-zero Chern number only if $\mu'<2$.

\begin{figure}
\includegraphics[width=0.8\linewidth]{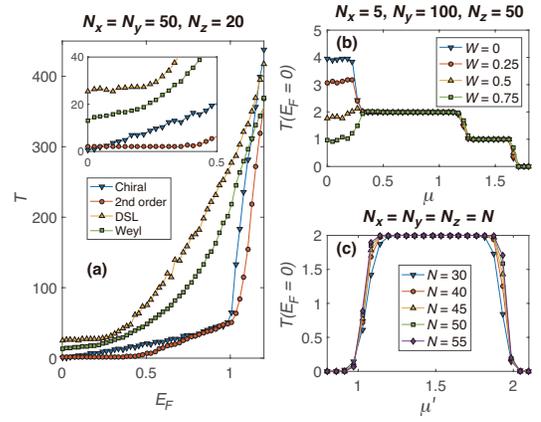}
\caption{(a) Conductance $T$ as a function of Fermi energy $E_F$ for the four topological phases created by our general approach. Inset zooms in to $0<E_F<0.5$. (b) $T$ versus $\mu$ at $E_F=0$ for a disordered sample with different strengths $W$, with the gap term given by Eq.~(\ref{Hgap_SNL}) and $m_0=0.8$, averaged over 100 disorder realizations. For small $\mu'$,  the system is an insulator with DSL and $T$ is not robust. For larger $\mu'$, the system is a Weyl semimetal and the conductance is robust and quantized due to
edge state conductance on the Fermi arc.  (c) $T$ versus $\mu'$ at $E_F=0$, for a second-order topological insulator using the gap term defined in Eq.~(\ref{gap_HO4}), with parameters $m=2,m_0=1.5$. The transitions at $\mu'=1$ and $\mu'=2$ agree with our theory. }\label{fig5}
\end{figure}

{\it Transport properties.}
To conclude our study, we now examine how the surface states in each class of topological phase may lead to different transport properties. We consider a 3D sample of our model with metallic leads (in the wide band limit)  attached along the $y$ direction, and computationally investigate the differential conductance $T$ (in units of $e^2/h$) using the Caroli formula {\cite{Caroli,sup}} 
. In Fig. \ref{fig5}(a) we illustrate $T$ versus Fermi energy $E_F$ for the four topological phases engineered from our general approach. For the chiral topological insulator, as $E_F$ increases, the Fermi surfaces evolve from a point at $E_F=0$ to loops of sizes proportional to $E_F$. Beyond a transition point at $E_F\sim 1$, one eventually arrives at the bulk bands and expects much larger conductance.  The corresponding conductance behavior agrees with these insights.
For the Weyl semimetal phase, the system is always metallic for finite $E_F$, which is reflected by the steady increase of the conductance. For the DSL insulator, it is curious to see that the conductance first stays around a certain value. This is because as $E_F$ increases, the Fermi surface intersects with two Dirac cones, resulting in two 1D loops, one increasing in size and the other decreasing in size.  As $E_F$ continues to increase, the surface states merge into the bulk ($E_F\sim0.25$), where the system becomes metallic and the conductance increases at a much faster rate.  For the second-order topological insulator, we observe two transitions in the conductance. For $E_F\lesssim0.4$, the plateau of $T$ is due to the 1D surface states along the hinges, which are similar to the surface states of a 2D Chern insulator. For $0.4\lesssim E_F\lesssim1$, gapped surface states start to contribute and conductance increases with $E_F$. Finally, $E_F\gtrsim1$ corresponds to bulk spectrum and the conductance also becomes that of a metal.
Like in other cases, these transition values of $E_F$ are in good agreement with our numerical results based on the system spectrum alone.

Since the DSL insulator and Weyl semimetal phases share similar conductance behavior as shown in Fig.~\ref{fig5}(a), we consider the responses of these two topological phases to disorder so as to distinguish between them. In Fig.~\ref{fig5}(b), we look into the conductance $T$ as a function of $\mu$, for a sample with on-site disorder, diagonal in both $s$ and $\tau$ space, uniformly distributed in $[-W,W]$.  We set $m_0=0.8$ and the Fermi energy $E_F=0$.
As discussed above, tuning $\mu$ induces a transition from a DSL insulator to a Weyl semimetal.  Consistent with this,
the Weyl semimetal for $(0.4\lesssim\mu\lesssim 1.5)$ exhibits robust quantized plateaux due to the quantum anomalous Hall effect along the Fermi arc connecting the nodal points \cite{Takane}. By contrast, the DSL phase is not resistant against disorder, as reflected by the curves in the leftmost part of Fig.~\ref{fig5}(b). This is justified since the disorder we apply is diagonal in both $s$ and $\tau$ space, and the presence of $s_0\tau_0$ violates the symmetry (\ref{eq:DSLsymmetry}) needed to protect the DSL.

Finally,  we verify the criterion for the existence of second-order topological insulator. We set $m=2$ as in Fig.~\ref{fig4} for the gap term given by Eq. (\ref{gap_HO4}). Fig.~\ref{fig5}(c) presents the conductance at $E_F=0$ as a function of $\mu'$.  The plateau due to the hinge states is clearly observed. The phase transition points are identified at $\mu'=1$ and $\mu'=2$, in agreement with our previous discussion. The transition becomes more pronounced as the sample size $N$ increases.

{\it Summary.}  We have proposed a rather general approach to the generation of a rich variety of topological phases. This task is made straightforward, at least in theory, by {\it directly} engineering the dispersion relations of their topological surface states.
In particular, a semimetal phase with a single 3D nodal loop is subject to a gap term of various forms of 2D spin-orbit coupling.
Under certain conditions, the 3D nodal-loop semimetal phase alone fully controls the existence of surface states, and the gap term we introduce takes complete charge of the dispersion relation of surface states. Not only is this insight expected to be useful in generating many interesting phases, we also highlight four intriguing phases in this work: (i) a chiral topological insulator with two-fold degenerate surface Dirac cones, (ii) an insulator with two-fold degenerate surface loop, (iii) a Weyl semimetal with four disconnected Fermi arcs, and (iv) a second-order topological insulator with 1D surface states along the hinges of a 3D system.  When necessary,
we have also discussed the symmetry protection and topological invariants of these phases.  Two-terminal transport calculations complement and confirm our analyses, offering a possibility to distinguish between these four topological phases.

{\it Acknowledgments.}  J.G. is supported by the Singapore NRF grant No. NRF-NRFI2017-04 (WBS No. R-144-000-378-281) and by the Singapore Ministry of Education Academic Research Fund Tier I (WBS No. R-144-000-353-112).
MANA acknowledges  partial  support  from  FCT  through grant  UID/CTM/04540/2013.   We thank Dr. Ching Hua Lee for interesting discussions during the completion of this work.

\appendix
\vspace{1cm}

\begin{center}{\bf Appendix} \end{center}
This appendix has two sections. In Sec.~A, we present theoretical details regarding the conditions for topological surface states to exist and how their dispersion relations can be obtained based on a gap term alone.  Sec.~B contains some materials describing our two-terminal transport calculations.

\section{Predictions of surface states based on effective Hamiltonians}
\subsection{Existence of surface states}
Given the effective Hamiltonian
\begin{eqnarray}
H_{\mathrm{eff},\pm}&=&\left(\sum_{i=x,y,z}\cos{k_i}-m\right)\tau_1+\sin{k_z}\tau_3\nonumber\\
&&+\left(M_0(\mathbf{k}_{\parallel})\pm\sqrt{\sum_iM_i^2(\mathbf{k}_{\parallel})}\right)\tau_2{\stackrel{\mathrm{def}}{=}\mathbf{h}_\pm\cdot\bm{\tau} }\label{Heff},
\end{eqnarray}
to study the existence of edge states localized along $z$, we consider the real-space version of (\ref{Heff}) in a semi-infinite geometry, terminated at the $x$--$y$ plane at $z=0$. Hence translational symmetry is broken in the $z$ direction, and surface states localized along $z$ may emerge. We assume that translational symmetry is preserved along $x$ and $y$, so that the corresponding momenta $k_x$ and $k_y$ are good quantum numbers. In this way, a higher dimensional system can be decoupled into a family of 1D problems parameterized by $k_x$ and $k_y$.

The three Pauli matrices form a Clifford algebra and span a 3-dimensional vector space, whereas $\mathbf{h}_{\pm}$ denote some vectors in this space. For a 1D problem with fixed $k_x$ and $k_y$, $\mathbf{h}_{\pm}$ is a function of $k_z$, and traces out a closed trajectory in the 3D vector space as $k_z$ varies from $0$ to $2\pi$. It was shown (Ref. \cite{Mong_2011}, section II and III) that the existence of surface states and their eigenenergies are determined by the position of the trajectory. To be specific, one can find a plane that contains the trajectory, and project the origin onto this plane. Whether the trajectory encloses the projected point determines the existence of surface states, and the distance between the origin and the plane gives the absolute value of the corresponding eigenenergies.

In our model, the trajectory is in the $1$--$3$ plane as $k_z$ only appears in the coefficients of $\tau_1$ and $\tau_3$. Thus we can write the Hamiltonian as $\mathbf{h}_{\pm}=\mathbf{h}^0_{\pm}+\mathbf{h}_{\pm,\parallel}+\mathbf{h}_{\pm,\perp}$, where
\begin{eqnarray}
\mathbf{h}^0_{\pm}&=&(\cos{k_z},0,\sin{k_z}),\nonumber\\
\mathbf{h}_{\pm,\parallel}&=&(\cos{k_x}+\cos{k_y}-m,0,0),\nonumber\\
\mathbf{h}_{\pm,\perp}&=&\left(0,\left(M_0(\mathbf{k}_{\parallel})\pm\sqrt{\sum_iM_i^2(\mathbf{k}_{\parallel})}\right),0\right).
\end{eqnarray}
Here $\mathbf{h}^0_{\pm}$ describes the trajectory which is a circle, while $\mathbf{h}_{\pm,\parallel}$ and $\mathbf{h}_{\pm,\perp}$ are $k_z$-independent, which lies in and normal to the $1$--$3$ plane respectively. The circular trajectory now has a radius of $1$, that is to say the projection of the origin on the $1$--$3$ plane is within the trajectory as long as
\begin{eqnarray}
|\mathbf{h}_{\pm,\parallel}|=|\cos{k_x}+\cos{k_y}-m|<1,\label{condition_1}
\end{eqnarray}
which is the condition for the existence of surface states.  Provided that Eq.(\ref{condition_1}) holds, the energies of the surface states are given by
\begin{eqnarray}
\epsilon_{u,\pm}=u|\mathbf{h}_{\pm,\perp}^0|=u\left| M_0(\mathbf{k}_{\parallel})\pm\sqrt{\sum_iM_i^2(\mathbf{k}_{\parallel}})\right|,\label{condition_2}
\end{eqnarray}
with $u=\pm1$ indicating the upper and lower branches of the edge states.

Since we consider gap terms that do not depend on $k_z$ and anticommute with the original nodal loop Hamiltonian $H_0$, we see that the existence of surface states is only associated with $H_0$, and the gap terms will modify the energies of these surface states.

\subsection{Surface state dispersion relations based on transfer matrix analysis}

For convenience, we rotate the pseudospin with an angle of $\pi/2$ in the $2$--$3$ plane, thus $\tau_2\rightarrow\tau_3$ and $\tau_3\rightarrow -\tau_2$, and rewrite the effective Hamiltonian as
\begin{eqnarray}
H_{\mathrm{eff},\pm}(k_z)=(\Delta_1+\cos{k_z})\tau_1-\sin{k_z}\tau_2+\Delta_2^{\pm}\tau_3,
\end{eqnarray}
with $\Delta_1=\cos{k_x}+\cos{k_y}-m$ and $\Delta_2^{\pm}=\left(M_0(\mathbf{k}_{\parallel})\pm\sqrt{\sum_iM_i^2(\mathbf{k}_{\parallel})}\right)$.
In real space, the Hamiltonian reads
\begin{eqnarray}
H_{\mathrm{eff},\pm}&=&\sum_{z}\Delta_1\hat{a}^{\dagger}_{z}\hat{b}_{z}+\hat{a}^{\dagger}_{z}\hat{b}_{z+1}+h.c.\nonumber\\
&&+\sum_{z}\Delta_{2}^{\pm}(\hat{a}^{\dagger}_{z}\hat{a}_{z}-\hat{b}^{\dagger}_{z}\hat{b}_{z}).
\end{eqnarray}
We consider a semi-infinite system with $z=1,2,3,...$, and choose the wavefunction as $\Psi_{n,\pm}=\sum_{z}(\psi_{a,z}\hat{a}_z^{\dagger}+\psi_{b,z}\hat{b}_z^{\dagger})|0\rangle$, with $|0\rangle$ the vacuum state. The eigenvalue equation $H_{\mathrm{eff},\pm}\Psi_{n,\pm}=E_{n,\pm}\Psi_{n,\pm}$ yields

\begin{eqnarray}
\Delta_1\psi_{b,z}+\psi_{b,z+1}+\Delta_2^{\pm}\psi_{a,z}=E_n\psi_{a,z}\label{eigenEq_a},\\
\Delta_1\psi_{a,z}+\psi_{a,z-1}-\Delta_2^{\pm}\psi_{b,z}=E_n\psi_{b,z}.\label{eigenEq_b}
\end{eqnarray}
Since $z$ starts from $1$, one has $\psi_{a,0}=0$ for the boundary condition, and $\psi_{b,0}$ is irrelevant here because it does not appear in Eqs.~(\ref{eigenEq_a}) and (\ref{eigenEq_b}).

A solution localized near $z=1$ can be achieved by choosing $E_n=-\Delta_2^{\pm}$. In such case, Eq. (\ref{eigenEq_b}) becomes
\begin{eqnarray}
\Delta_1\psi_{a,z}+\psi_{a,z-1}&=&0, \label{eigenEq_d}
\end{eqnarray}
and we have $\psi_{a,z}=0$ for any $z$ due to the boundary condition. Thus Eq. (\ref{eigenEq_a}) can also be simplified as
\begin{eqnarray}
\Delta_1\psi_{b,z}+\psi_{b,z+1}&=&0, \label{eigenEq_c}
\end{eqnarray}
which gives an exponentially decreasing state only if $|\Delta_1|<1$, whence Eq.~(\ref{condition_1}) as a criterion for the existence of edge states. Notice that here we choose a semi-infinite geometry with only one edge. When the system has a finite size with two ends, edge states with non-zero $a$ component localized at the other end can exist when $E_n=\Delta_2^{\pm}$.

Next we consider the case with general eigenenergy $E_n\neq \pm \Delta_2^{\pm}$. Substituting Eq. (\ref{eigenEq_a}) to (\ref{eigenEq_b}), we have
\begin{eqnarray}
\frac{\Delta_1^2+1}{\Delta_1}\psi_{b,z}+\psi_{b,z+1}+\psi_{b,z-1}=\frac{E_n^2-(\Delta_2^{\pm})^2}{\Delta_1}\psi_{b,z}.
\end{eqnarray}
In the form of transfer matrix, we have
\begin{eqnarray}
\left(\begin{array}{c}
\psi_{b,z+1}\\
\psi_{b,z}
\end{array}
\right)=
A
\left(\begin{array}{c}
\psi_{b,z}\\
\psi_{b,z-1}
\end{array}
\right),~~~A=\left(\begin{array}{cc}
\frac{E_n^2-(\Delta_2^{\pm})^2-\Delta_1^2-1}{\Delta_1} & -1\\
1 & 0
\end{array}
\right).
\end{eqnarray}
We now apply the method in Ref. \cite{SANCHEZSOTO2012,Li2015} to determine whether the eigenstate corresponding to $E_n$ is an edge state or not. Assuming
\begin{eqnarray}
A\left(\begin{array}{c}
a\\
b
\end{array}
\right)=\epsilon_A\left(\begin{array}{c}
a\\
b
\end{array}\right),\label{trans}
\end{eqnarray}
 the vector $(\psi_{b,z+1}, \psi_{b,z})^T$ can be written as a linear superposition of the two orthogonal eigenvectors of $A$, i.e., $(\psi_{b,z+1}, \psi_{b,z})^T = s_1(a_1, b_1)^T + s_2(a_2, b_2)^T$, with $s_1$ and $s_2$ the superposition coefficients. Thus we have
\begin{align}
\left(\begin{array}{c}
\psi_{b,z+n+1}\\
\psi_{b,z+n}
\end{array}
\right)&=
A^n
\left(\begin{array}{c}
\psi_{b,z+1}\\
\psi_{b,z}
\end{array}
\right)  =s_1\epsilon_{A,1}^n\left(\begin{array}{c}
a_1\\
b_1
\end{array}\right)+s_2\epsilon_{A,2}^n\left(\begin{array}{c}
a_2\\
b_2
\end{array}\right),
\end{align}
which suggests that $\psi_z=0$ as $z\rightarrow \infty$ if both eigenvalues of $A$ are less than unity. On the other hand, if both eigenvalues are greater than unity, it corresponds to an edge state localized at $z\rightarrow \infty$.

In our case, the eigenvalues of the transfer matrix $A$ in Eq. (\ref{trans}) satisfy
\begin{eqnarray}
\epsilon_{A,1}\epsilon_{A,2}=\det{A}=1.
\end{eqnarray}
Hence the two eigenvalues cannot be both greater or less than unity at the same time. That is to say the wavefunction of $E_n\neq \pm\Delta_2^{\pm}$ cannot localize anywhere along the 1D chain.

As a conclusion, the existence of edge states (or surface states as in the main text) requires
\begin{eqnarray}
|\Delta_1|=|\cos{k_x}+\sin{k_y}-m|<1,
\end{eqnarray}
and their energies are given by
\begin{eqnarray}
E=\pm\Delta_2^{\pm}=\pm\left(M_0(\mathbf{k}_{\parallel})\pm\sqrt{\sum_iM_i^2(\mathbf{k}_{\parallel})}\right).
\end{eqnarray}

\section{Details of two-terminal transport calculations}

To distingiush between the different types of surface states, we presented in the main text
their respective transport properties.  We consider a 3D sample of our model with metallic leads attached along the $y$ direction, and calculate the differential conductance $T$ (in units of $e^2/h$) using the Caroli formula \cite{Caroli}:  $T(E_F)=\mathrm{Tr}[G\Gamma_RG^\dag \Gamma_L]$. Here $G=(E_F-H-\Sigma_L-\Sigma_R)^{-1}$ is the retarded Green's function, $\Sigma_{L/R}$ is the self energy due to the left (right) lead (taken in the wideband limit $\Sigma_{L/R}=-i\delta_{ss'}\delta_{\tau\tau'}\delta_{ij}/2$ for simplicity), $\Gamma_{L/R}=i(\Sigma_{L/R}-\Sigma^\dag_{L/R})$ is the line-width function, and $\mathrm{Tr}[\dots]$ is a trace over both the spin and pseudospin subspaces. For efficient simulations of large systems we employ the method of recursive Green's functions \cite{Recursive} which reduces the complexity from $O(N_x^3N_y^3N_z^3)$ to $O(N_x^3N_yN_z^3)$.


\bibliographystyle{apsrev4-1}

\begin{thebibliography}{43}
\bibitem [{not()}]{note}%
  \BibitemOpen
  \href@noop {} {\ }\bibinfo {note} {The sketches are only for small $m$ and
  $m_0$, so that $\mathbf{L}_0$ and the SNL are both nearly perfect circles. In
  general, the zeros of $H_{gap}$ give four closed loops in the surface
  Brillouin zone when $\mu=0$. For larger $m_0$, each two of these loops will
  merge into one loop when increasing $\mu$. Nevertheless, the topology and the
  structure of surface states are hardly affected, and the transition from a
  SNL insulator to a trivial band insulator through a nodal point semimetallic
  phase still exists.}\BibitemShut {Stop}%
\end{thebibliography}

\end{document}